\documentclass{article}[10pt,letterpaper]
\usepackage{fancyhdr}
\usepackage{supertabular}
\usepackage{color}
\usepackage{amsmath}
\usepackage{pbox}
\usepackage{amssymb}
\usepackage{amsbsy}
\usepackage{scalefnt}
\usepackage{ifpdf}
\usepackage{graphicx}
\usepackage{multicol}
\usepackage[colorlinks=true, linkcolor=blue, linktocpage=true,urlcolor=blue]{hyperref}
\usepackage{longtable}
\usepackage{needspace}
\usepackage{eso-pic}
\usepackage{mathptmx}
\usepackage{multirow}
\usepackage{lscape}
\usepackage{textcomp}
\usepackage{changepage}
\usepackage{xfrac}
\usepackage{txfonts}
\usepackage{pdfcomment}
\usepackage[absolute]{textpos}
\ifpdf
\fi
\makeatletter
\renewcommand{\l@section}{\@dottedtocline{2}{1em}{0em}}
\renewcommand{\l@subsection}{\@dottedtocline{3}{2.5cm}{0em}}
\renewcommand{\hrulefill}{\leavevmode \leaders \hrule \@height 1pt \hfill \kern\z@}
\makeatother
\renewcommand{\underline}[1]{\begin{tabular}{@{\extracolsep{\fill}}c@{\extracolsep{\fill}}}#1\\[-0.2cm]\hrulefill\end{tabular}}

\newcounter{chappage}
\AddToShipoutPicture{\stepcounter{chappage}}
\fancypagestyle{plain}{\lhead{}\rhead{}}
\fancypagestyle{single}{\lhead{\leftmark \ \\}\rhead{\leftmark \ \\}}
\fancypagestyle{bob}{\lhead{\leftmark -\thechappage \ \\}\rhead{\leftmark -\thechappage \ \\}}
\begin{document}
\fontsize{9}{10}\selectfont
\fontdimen2\font=1.3\fontdimen2\font
\thispagestyle{empty}
\setcounter{page}{1}
\begin{center}

\vspace{0.5cm}
{ \huge Energy Levels of \ensuremath{^{\textnormal{15}}}Be*}\\
\vspace{1.0cm}
{ \normalsize K. Setoodehnia\ensuremath{^{\textnormal{1,2}}} and J. H. Kelley\ensuremath{^{\textnormal{1,3}}}}\\
\vspace{0.2in}
{ \small \it \ensuremath{^{\textnormal{1}}}Triangle Universities Nuclear Laboratory, Duke University,\\
  Durham, North Carolina 27708, USA.\\
  \ensuremath{^{\textnormal{2}}}Department of Physics, Duke University, Durham, North Carolina\\
  27708, USA.\\
  \ensuremath{^{\textnormal{3}}}Department of Physics, North Carolina State University,\\
  Raleigh, North Carolina 27607, USA}\\
\vspace{0.2in}
\end{center}

\setlength{\parindent}{-0.5cm}
\addtolength{\leftskip}{2cm}
\addtolength{\rightskip}{2cm}
{\bf Abstract: }
In this document, experimental nuclear structure data are evaluated for \ensuremath{^{\textnormal{15}}}Be. The details of each reaction populating \ensuremath{^{\textnormal{15}}}Be levels are compiled and evaluated. The combined results provide a set of adopted values that include level energies, spins and parities, level widths, decay types and other nuclear properties.\\

{\bf Cutoff Date: }
Literature available up to January 31, 2025 has been considered; the primary bibliographic source, the NSR database (\href{https://www.nndc.bnl.gov/nsr/nsrlink.jsp?2011Pr03,B}{2011Pr03}) available at Brookhaven National Laboratory web page: www.nndc.bnl.gov/nsr/.\\

{\bf General Policies and Organization of Material: }
See the April 2025 issue of the {\it Nuclear Data Sheets} or \\https://www.nndc.bnl.gov/nds/docs/NDSPolicies.pdf. \\

{\bf Acknowledgements: }
The authors expresses her gratitude to personnel at the National Nuclear Data Center (NNDC) at Brookhaven National Laboratory for facilitating this work.\\

\vfill

* This work is supported by the Office of Nuclear Physics, Office of Science, U.S. Department of Energy under contracts: DE-FG02-97ER41042 {\textminus} North Carolina State University and DE-FG02-97ER41033 {\textminus} Duke University\\

\setlength{\parindent}{+0.5cm}
\addtolength{\leftskip}{-2cm}
\addtolength{\rightskip}{-2cm}
\newpage
\pagestyle{plain}
\setlength{\columnseprule}{1pt}
\setlength{\columnsep}{1cm}
\begin{center}
\underline{\normalsize Index for A=15}
\end{center}
\hspace{.3cm}\raggedright\underline{Nuclide}\hspace{1cm}\underline{Data Type\mbox{\hspace{2.3cm}}}\hspace{2cm}\underline{Page}\hspace{1cm}
\raggedright\underline{Nuclide}\hspace{1cm}\underline{Data Type\mbox{\hspace{2.3cm}}}\hspace{2cm}\underline{Page}
\begin{adjustwidth}{}{0.05\textwidth}
\begin{multicols}{2}
\setcounter{tocdepth}{3}
\renewcommand{\contentsname}{\protect\vspace{-0.8cm}}
\tableofcontents
\end{multicols}
\end{adjustwidth}
\clearpage
\thispagestyle{empty}
\mbox{}
\clearpage
\clearpage
\pagestyle{bob}
\begin{center}
\section[\ensuremath{^{15}_{\ 4}}Be\ensuremath{_{11}^{~}}]{ }
\vspace{-30pt}
\setcounter{chappage}{1}
\subsection[\hspace{-0.2cm}Adopted Levels]{ }
\vspace{-20pt}
\vspace{0.3cm}
\hypertarget{BE0}{{\bf \small \underline{Adopted \hyperlink{15BE_LEVEL}{Levels}}}}\\
\vspace{4pt}
\vspace{8pt}
\parbox[b][0.3cm]{17.7cm}{\addtolength{\parindent}{-0.2in}Q(\ensuremath{\beta^-})=20665 {\it 29}; S(n)=$-$1.60\ensuremath{\times10^{3}} {\it 13}\hspace{0.2in}\href{https://www.nndc.bnl.gov/nsr/nsrlink.jsp?2021Wa16,B}{2021Wa16},\href{https://www.nndc.bnl.gov/nsr/nsrlink.jsp?2024Ku30,B}{2024Ku30}}\\
\parbox[b][0.3cm]{17.7cm}{\addtolength{\parindent}{-0.2in}Q(\ensuremath{\beta}\ensuremath{^{-}}): Computed using \ensuremath{\Delta}M=49622 keV \textit{20}, which is deduced using S\ensuremath{_{\textnormal{3n}}}(\ensuremath{^{\textnormal{15}}}Be\ensuremath{_{\textnormal{g.s.}}})={\textminus}330 keV \textit{20} (\href{https://www.nndc.bnl.gov/nsr/nsrlink.jsp?2024Ku30,B}{2024Ku30}) and the mass excesses}\\
\parbox[b][0.3cm]{17.7cm}{of \ensuremath{^{\textnormal{12}}}Be and the neutron from (\href{https://www.nndc.bnl.gov/nsr/nsrlink.jsp?2021Wa16,B}{2021Wa16}: AME-2020). This \ensuremath{\Delta}M value compares with \ensuremath{\Delta}M=49826 keV \textit{166} (\href{https://www.nndc.bnl.gov/nsr/nsrlink.jsp?2021Wa16,B}{2021Wa16}), which}\\
\parbox[b][0.3cm]{17.7cm}{is determined using the result of (\href{https://www.nndc.bnl.gov/nsr/nsrlink.jsp?2013Sn02,B}{2013Sn02}), where a state in \ensuremath{^{\textnormal{15}}}Be was first observed with S\ensuremath{_{\textnormal{n}}}(\ensuremath{^{\textnormal{14}}}Be+n)={\textminus}1.8 MeV \textit{1}; this state is}\\
\parbox[b][0.3cm]{17.7cm}{now identified as the first excited state.}\\
\parbox[b][0.3cm]{17.7cm}{\addtolength{\parindent}{-0.2in}S(n): From E\ensuremath{_{\textnormal{c.m.}}}(n+\ensuremath{^{\textnormal{14}}}Be)=1600 keV \textit{132} for \ensuremath{^{\textnormal{15}}}Be\ensuremath{_{\textnormal{g.s.}}}. This is obtained using E\ensuremath{_{\textnormal{c.m.}}}(3n+\ensuremath{^{\textnormal{12}}}Be\ensuremath{_{\textnormal{g.s.}}})=330 keV \textit{20} (\href{https://www.nndc.bnl.gov/nsr/nsrlink.jsp?2024Ku30,B}{2024Ku30}), and n,}\\
\parbox[b][0.3cm]{17.7cm}{\ensuremath{^{\textnormal{12}}}Be and \ensuremath{^{\textnormal{14}}}Be masses from (\href{https://www.nndc.bnl.gov/nsr/nsrlink.jsp?2021Wa16,B}{2021Wa16}). Note E\ensuremath{_{\textnormal{c.m.}}}(n+\ensuremath{^{\textnormal{14}}}Be)=E\ensuremath{_{\textnormal{c.m.}}}(3n+\ensuremath{^{\textnormal{12}}}Be)+S\ensuremath{_{\textnormal{2n}}}(\ensuremath{^{\textnormal{14}}}Be) and S\ensuremath{_{\textnormal{2n}}}(\ensuremath{^{\textnormal{14}}}Be)=1270 keV \textit{130}}\\
\parbox[b][0.3cm]{17.7cm}{(\href{https://www.nndc.bnl.gov/nsr/nsrlink.jsp?2021Wa16,B}{2021Wa16}: AME-2020).}\\

\vspace{0.385cm}
\parbox[b][0.3cm]{17.7cm}{\addtolength{\parindent}{-0.2in}\textit{Theoretical works (see also the \ensuremath{^{2}}H(\ensuremath{^{\textnormal{14}}}Be,\ensuremath{^{\textnormal{15}}}Be) dataset)}:}\\
\parbox[b][0.3cm]{17.7cm}{\addtolength{\parindent}{-0.2in}\href{https://www.nndc.bnl.gov/nsr/nsrlink.jsp?2006Ko02,B}{2006Ko02}: A chiral soliton model with a \ensuremath{\approx} 30\% rescaling of the Skyrme constant is used to estimate the mass excess of A=6 to}\\
\parbox[b][0.3cm]{17.7cm}{32 nuclides. Also see calculations in (\href{https://www.nndc.bnl.gov/nsr/nsrlink.jsp?1987Sa15,B}{1987Sa15}, \href{https://www.nndc.bnl.gov/nsr/nsrlink.jsp?1985Po10,B}{1985Po10}, \href{https://www.nndc.bnl.gov/nsr/nsrlink.jsp?1981Se06,B}{1981Se06}).}\\
\parbox[b][0.3cm]{17.7cm}{\addtolength{\parindent}{-0.2in}\href{https://www.nndc.bnl.gov/nsr/nsrlink.jsp?2018Du01,B}{2018Du01}: Calculated the energy and width of the ground state (resonance state) of \ensuremath{^{\textnormal{15}}}Be using a density dependent M3Y}\\
\parbox[b][0.3cm]{17.7cm}{microscopic potential.}\\
\parbox[b][0.3cm]{17.7cm}{\addtolength{\parindent}{-0.2in}\href{https://www.nndc.bnl.gov/nsr/nsrlink.jsp?2018Fo22,B}{2018Fo22}: Calculated spectroscopic factors and single-particle widths for the neutron decay of the \ensuremath{^{\textnormal{15}}}Be*(5/2\ensuremath{^{\textnormal{+}}_{\textnormal{1}}}) state to the}\\
\parbox[b][0.3cm]{17.7cm}{\ensuremath{^{\textnormal{14}}}Be(g.s.) and \ensuremath{^{\textnormal{14}}}Be*(2\ensuremath{^{\textnormal{+}}_{\textnormal{1}}}) state using a simple model. Results are compared with the experimental width for the \ensuremath{^{\textnormal{15}}}Be*(5/2\ensuremath{^{\textnormal{+}}_{\textnormal{1}}})}\\
\parbox[b][0.3cm]{17.7cm}{state from (\href{https://www.nndc.bnl.gov/nsr/nsrlink.jsp?2013Sn02,B}{2013Sn02}). This state was initially assumed to be the ground state by (\href{https://www.nndc.bnl.gov/nsr/nsrlink.jsp?2013Sn02,B}{2013Sn02}, \href{https://www.nndc.bnl.gov/nsr/nsrlink.jsp?2018Fo22,B}{2018Fo22}).}\\
\parbox[b][0.3cm]{17.7cm}{\addtolength{\parindent}{-0.2in}\href{https://www.nndc.bnl.gov/nsr/nsrlink.jsp?2019Fo09,B}{2019Fo09}: Investigated the decay width for the \ensuremath{^{\textnormal{16}}}Be\ensuremath{\rightarrow}\ensuremath{^{\textnormal{14}}}Be+2n decay via a 2n cluster decay and a sequential decay through the}\\
\parbox[b][0.3cm]{17.7cm}{\ensuremath{^{\textnormal{15}}}Be*(5/2\ensuremath{^{\textnormal{+}}_{\textnormal{1}}}) state (which was interpreted as the ground state) or through a hypothetical J\ensuremath{^{\ensuremath{\pi}}}=1/2\ensuremath{^{\textnormal{+}}} resonance in \ensuremath{^{\textnormal{15}}}Be. Comparison}\\
\parbox[b][0.3cm]{17.7cm}{with experimental \ensuremath{^{\textnormal{16}}}Be decay width is discussed.}\\
\parbox[b][0.3cm]{17.7cm}{\addtolength{\parindent}{-0.2in}\href{https://www.nndc.bnl.gov/nsr/nsrlink.jsp?2020Wi12,B}{2020Wi12}: Analyzed available data for various nuclei, including \ensuremath{^{\textnormal{15}}}Be; deduced the appearance and disappearance of shell closures}\\
\parbox[b][0.3cm]{17.7cm}{at N=Z=8, 14, 16, and 20.}\\
\vspace{12pt}
\hypertarget{15BE_LEVEL}{\underline{$^{15}$Be Levels}}\\
\begin{longtable}[c]{ll}
\multicolumn{2}{c}{\underline{Cross Reference (XREF) Flags}}\\
 \\
\hyperlink{BE1}{\texttt{A }}& \ensuremath{^{\textnormal{2}}}H(\ensuremath{^{\textnormal{14}}}Be,\ensuremath{^{\textnormal{15}}}Be)\\
\hyperlink{BE2}{\texttt{B }}& \ensuremath{^{\textnormal{9}}}Be(\ensuremath{^{\textnormal{17}}}C,\ensuremath{^{\textnormal{15}}}Be2p)\\
\hyperlink{BE3}{\texttt{C }}& \ensuremath{^{\textnormal{12}}}C(\ensuremath{^{\textnormal{18}}}C,\ensuremath{^{\textnormal{15}}}Be)\\
\end{longtable}
\vspace{-0.5cm}
\begin{longtable}{ccccccccc@{\extracolsep{\fill}}c}
\multicolumn{2}{c}{E(level)$^{{\hyperlink{BE0LEVEL0}{a}}}$}&J$^{\pi}$$^{}$&\multicolumn{2}{c}{\ensuremath{\Gamma}$^{}$}&\multicolumn{2}{c}{E\ensuremath{_{\textnormal{c.m.}}}(\ensuremath{^{\textnormal{14}}}Be\ensuremath{_{\textnormal{g.s.}}}+n) (keV)$^{}$}&XREF&Comments&\\[-.2cm]
\multicolumn{2}{c}{\hrulefill}&\hrulefill&\multicolumn{2}{c}{\hrulefill}&\multicolumn{2}{c}{\hrulefill}&\hrulefill&\hrulefill&
\endfirsthead
\multicolumn{1}{r@{}}{0}&\multicolumn{1}{@{}l}{}&\multicolumn{1}{l}{(3/2\ensuremath{^{+}})}&\multicolumn{1}{r@{}}{$<$200}&\multicolumn{1}{@{}l}{}&\multicolumn{1}{r@{}}{1.60\ensuremath{\times10^{3}}}&\multicolumn{1}{@{ }l}{{\it 13}}&\multicolumn{1}{l}{\texttt{\hyperlink{BE1}{A}\ \ } }&\parbox[t][0.3cm]{6.71537cm}{\raggedright \%n\ensuremath{\approx}100 (\href{https://www.nndc.bnl.gov/nsr/nsrlink.jsp?2024Ku30,B}{2024Ku30})\vspace{0.1cm}}&\\
&&&&&&&&\parbox[t][0.3cm]{6.71537cm}{\raggedright \ensuremath{\Gamma}: From (\href{https://www.nndc.bnl.gov/nsr/nsrlink.jsp?2024Ku30,B}{2024Ku30}): The best fit resulted in\vspace{0.1cm}}&\\
&&&&&&&&\parbox[t][0.3cm]{6.71537cm}{\raggedright {\ }{\ }{\ }an optimal \ensuremath{\Gamma}=110 keV.\vspace{0.1cm}}&\\
&&&&&&&&\parbox[t][0.3cm]{6.71537cm}{\raggedright J\ensuremath{^{\pi}}: From (\href{https://www.nndc.bnl.gov/nsr/nsrlink.jsp?2024Ku30,B}{2024Ku30}) and based on the shell\vspace{0.1cm}}&\\
&&&&&&&&\parbox[t][0.3cm]{6.71537cm}{\raggedright {\ }{\ }{\ }model predictions by (\href{https://www.nndc.bnl.gov/nsr/nsrlink.jsp?2011Sp01,B}{2011Sp01}) and\vspace{0.1cm}}&\\
&&&&&&&&\parbox[t][0.3cm]{6.71537cm}{\raggedright {\ }{\ }{\ }supported by the theoretical findings of\vspace{0.1cm}}&\\
&&&&&&&&\parbox[t][0.3cm]{6.71537cm}{\raggedright {\ }{\ }{\ }(\href{https://www.nndc.bnl.gov/nsr/nsrlink.jsp?2015Fo04,B}{2015Fo04}, \href{https://www.nndc.bnl.gov/nsr/nsrlink.jsp?2018Fo22,B}{2018Fo22}) regarding the decay\vspace{0.1cm}}&\\
&&&&&&&&\parbox[t][0.3cm]{6.71537cm}{\raggedright {\ }{\ }{\ }pattern of the 5/2\ensuremath{^{\textnormal{+}}_{\textnormal{1}}} state in \ensuremath{^{\textnormal{15}}}Be.\vspace{0.1cm}}&\\
&&&&&&&&\parbox[t][0.3cm]{6.71537cm}{\raggedright E\ensuremath{_{\textnormal{c.m.}}}(\ensuremath{^{\textnormal{14}}}Be\ensuremath{_{\textnormal{g.s.}}}+n) (keV): Deduced from adding\vspace{0.1cm}}&\\
&&&&&&&&\parbox[t][0.3cm]{6.71537cm}{\raggedright {\ }{\ }{\ }S\ensuremath{_{\textnormal{2n}}}(\ensuremath{^{\textnormal{14}}}Be)=1270 keV \textit{130} (\href{https://www.nndc.bnl.gov/nsr/nsrlink.jsp?2021Wa16,B}{2021Wa16}:\vspace{0.1cm}}&\\
&&&&&&&&\parbox[t][0.3cm]{6.71537cm}{\raggedright {\ }{\ }{\ }AME-2020) to E\ensuremath{_{\textnormal{c.m.}}}(3n+\ensuremath{^{\textnormal{12}}}Be)=330 keV \textit{20}\vspace{0.1cm}}&\\
&&&&&&&&\parbox[t][0.3cm]{6.71537cm}{\raggedright {\ }{\ }{\ }(\href{https://www.nndc.bnl.gov/nsr/nsrlink.jsp?2024Ku30,B}{2024Ku30}).\vspace{0.1cm}}&\\
&&&&&&&&\parbox[t][0.3cm]{6.71537cm}{\raggedright Decay mode: \ensuremath{^{\textnormal{14}}}Be*(2\ensuremath{^{\textnormal{+}}_{\textnormal{1}}})+n\ensuremath{\rightarrow}\ensuremath{^{\textnormal{12}}}Be\ensuremath{_{\textnormal{g.s.}}}+3n\vspace{0.1cm}}&\\
&&&&&&&&\parbox[t][0.3cm]{6.71537cm}{\raggedright {\ }{\ }{\ }(\href{https://www.nndc.bnl.gov/nsr/nsrlink.jsp?2024Ku30,B}{2024Ku30}).\vspace{0.1cm}}&\\
&&&&&&&&\parbox[t][0.3cm]{6.71537cm}{\raggedright As indicated in (\href{https://www.nndc.bnl.gov/nsr/nsrlink.jsp?2024Ku30,B}{2024Ku30}), the large \ensuremath{\Delta}E in\vspace{0.1cm}}&\\
&&&&&&&&\parbox[t][0.3cm]{6.71537cm}{\raggedright {\ }{\ }{\ }\ensuremath{^{\textnormal{15}}}Be\ensuremath{_{\textnormal{g.s.}}} is connected to the \ensuremath{\Delta}S\ensuremath{_{\textnormal{2n}}}(\ensuremath{^{\textnormal{14}}}Be)=130\vspace{0.1cm}}&\\
&&&&&&&&\parbox[t][0.3cm]{6.71537cm}{\raggedright {\ }{\ }{\ }keV. An improvement in the \ensuremath{^{\textnormal{14}}}Be mass would\vspace{0.1cm}}&\\
&&&&&&&&\parbox[t][0.3cm]{6.71537cm}{\raggedright {\ }{\ }{\ }resolve some level-energy uncertainties for the\vspace{0.1cm}}&\\
&&&&&&&&\parbox[t][0.3cm]{6.71537cm}{\raggedright {\ }{\ }{\ }levels decaying via \ensuremath{^{\textnormal{14}}}Be\ensuremath{_{\textnormal{g.s.}}}+n and those\vspace{0.1cm}}&\\
&&&&&&&&\parbox[t][0.3cm]{6.71537cm}{\raggedright {\ }{\ }{\ }decaying via \ensuremath{^{\textnormal{14}}}Be*(2\ensuremath{^{\textnormal{+}}_{\textnormal{1}}})+n\ensuremath{\rightarrow}\ensuremath{^{\textnormal{12}}}Be\ensuremath{_{\textnormal{g.s.}}}+3n.\vspace{0.1cm}}&\\
\multicolumn{1}{r@{}}{0.16\ensuremath{\times10^{3}}}&\multicolumn{1}{@{ }l}{{\it 16}}&\multicolumn{1}{l}{(5/2\ensuremath{^{+}})}&\multicolumn{1}{r@{}}{0}&\multicolumn{1}{@{.}l}{58 MeV {\it 20}}&\multicolumn{1}{r@{}}{1.76\ensuremath{\times10^{3}}}&\multicolumn{1}{@{ }l}{{\it 10}}&\multicolumn{1}{l}{\texttt{\hyperlink{BE1}{A}\ \hyperlink{BE3}{C}} }&\parbox[t][0.3cm]{6.71537cm}{\raggedright \%n\ensuremath{\approx}100 (\href{https://www.nndc.bnl.gov/nsr/nsrlink.jsp?2013Sn02,B}{2013Sn02},\href{https://www.nndc.bnl.gov/nsr/nsrlink.jsp?2021Co07,B}{2021Co07},\href{https://www.nndc.bnl.gov/nsr/nsrlink.jsp?2024Ku30,B}{2024Ku30})\vspace{0.1cm}}&\\
\end{longtable}
\begin{textblock}{29}(0,27.3)
Continued on next page (footnotes at end of table)
\end{textblock}
\clearpage
\begin{longtable}{cccccc@{\extracolsep{\fill}}c}
\\[-.4cm]
\multicolumn{7}{c}{{\bf \small \underline{Adopted \hyperlink{15BE_LEVEL}{Levels} (continued)}}}\\
\multicolumn{7}{c}{~}\\
\multicolumn{7}{c}{\underline{\ensuremath{^{15}}Be Levels (continued)}}\\
\multicolumn{7}{c}{~}\\
\multicolumn{2}{c}{E(level)$^{{\hyperlink{BE0LEVEL0}{a}}}$}&\multicolumn{2}{c}{E\ensuremath{_{\textnormal{c.m.}}}(\ensuremath{^{\textnormal{14}}}Be\ensuremath{_{\textnormal{g.s.}}}+n) (keV)$^{}$}&XREF&Comments&\\[-.2cm]
\multicolumn{2}{c}{\hrulefill}&\multicolumn{2}{c}{\hrulefill}&\hrulefill&\hrulefill&
\endhead
&&&&&\parbox[t][0.3cm]{11.07296cm}{\raggedright E\ensuremath{_{\textnormal{c.m.}}}(\ensuremath{^{\textnormal{14}}}Be\ensuremath{_{\textnormal{g.s.}}}+n) (keV): Weighted average of E\ensuremath{_{\textnormal{c.m.}}}(n+\ensuremath{^{\textnormal{14}}}Be)=1.8 MeV \textit{1}\vspace{0.1cm}}&\\
&&&&&\parbox[t][0.3cm]{11.07296cm}{\raggedright {\ }{\ }{\ }(\href{https://www.nndc.bnl.gov/nsr/nsrlink.jsp?2013Sn02,B}{2013Sn02}) and E\ensuremath{_{\textnormal{c.m.}}}(n+\ensuremath{^{\textnormal{14}}}Be)=1.70 MeV \textit{13} (\href{https://www.nndc.bnl.gov/nsr/nsrlink.jsp?2021Co07,B}{2021Co07}). These authors\vspace{0.1cm}}&\\
&&&&&\parbox[t][0.3cm]{11.07296cm}{\raggedright {\ }{\ }{\ }initially interpreted this state as the ground state.\vspace{0.1cm}}&\\
&&&&&\parbox[t][0.3cm]{11.07296cm}{\raggedright E(level): See also E\ensuremath{_{\textnormal{x}}}=210 keV, which is deduced from a measured value of\vspace{0.1cm}}&\\
&&&&&\parbox[t][0.3cm]{11.07296cm}{\raggedright {\ }{\ }{\ }E\ensuremath{_{\textnormal{c.m.}}}(3n+\ensuremath{^{\textnormal{12}}}Be)=540 keV (\href{https://www.nndc.bnl.gov/nsr/nsrlink.jsp?2024Ku30,B}{2024Ku30}). The latter value is equivalent to\vspace{0.1cm}}&\\
&&&&&\parbox[t][0.3cm]{11.07296cm}{\raggedright {\ }{\ }{\ }E\ensuremath{_{\textnormal{c.m.}}}(n+\ensuremath{^{\textnormal{14}}}Be)=540 keV+1270 keV=1810 keV.\vspace{0.1cm}}&\\
&&&&&\parbox[t][0.3cm]{11.07296cm}{\raggedright \ensuremath{\Gamma}: From \ensuremath{\Gamma}=575 keV \textit{200} (\href{https://www.nndc.bnl.gov/nsr/nsrlink.jsp?2013Sn02,B}{2013Sn02}).\vspace{0.1cm}}&\\
&&&&&\parbox[t][0.3cm]{11.07296cm}{\raggedright \ensuremath{\Gamma}: See also \ensuremath{\Gamma}\ensuremath{\geq}200 keV (\href{https://www.nndc.bnl.gov/nsr/nsrlink.jsp?2021Co07,B}{2021Co07}). (\href{https://www.nndc.bnl.gov/nsr/nsrlink.jsp?2015Fo04,B}{2015Fo04}) reported that \ensuremath{\Gamma}=0.58 MeV\vspace{0.1cm}}&\\
&&&&&\parbox[t][0.3cm]{11.07296cm}{\raggedright {\ }{\ }{\ }from (\href{https://www.nndc.bnl.gov/nsr/nsrlink.jsp?2013Sn02,B}{2013Sn02}) is larger than \ensuremath{\Gamma}=390 keV calculated using the shell model\vspace{0.1cm}}&\\
&&&&&\parbox[t][0.3cm]{11.07296cm}{\raggedright {\ }{\ }{\ }(\href{https://www.nndc.bnl.gov/nsr/nsrlink.jsp?2015Fo04,B}{2015Fo04}). They attributed the larger experimental width to another decay\vspace{0.1cm}}&\\
&&&&&\parbox[t][0.3cm]{11.07296cm}{\raggedright {\ }{\ }{\ }branch from the \ensuremath{^{\textnormal{15}}}Be*(5/2\ensuremath{^{\textnormal{+}}}) state to the \ensuremath{^{\textnormal{14}}}Be*(2\ensuremath{^{\textnormal{+}}_{\textnormal{1}}}) level.\vspace{0.1cm}}&\\
&&&&&\parbox[t][0.3cm]{11.07296cm}{\raggedright J\ensuremath{^{\pi}}: From (\href{https://www.nndc.bnl.gov/nsr/nsrlink.jsp?2013Sn02,B}{2013Sn02}): based on the decay patterns predicted by their shell-model\vspace{0.1cm}}&\\
&&&&&\parbox[t][0.3cm]{11.07296cm}{\raggedright {\ }{\ }{\ }calculations and the spectroscopic overlap of the proposed \ensuremath{^{\textnormal{15}}}Be levels with\vspace{0.1cm}}&\\
&&&&&\parbox[t][0.3cm]{11.07296cm}{\raggedright {\ }{\ }{\ }\ensuremath{^{\textnormal{14}}}Be\ensuremath{_{\textnormal{g.s.}}}, which were calculated by (\href{https://www.nndc.bnl.gov/nsr/nsrlink.jsp?2013Sn02,B}{2013Sn02}) using FRESCO: S=0.44 if\vspace{0.1cm}}&\\
&&&&&\parbox[t][0.3cm]{11.07296cm}{\raggedright {\ }{\ }{\ }J\ensuremath{^{\ensuremath{\pi}}}=5/2\ensuremath{^{\textnormal{+}}} and S=0.07 if J\ensuremath{^{\ensuremath{\pi}}}=3/2\ensuremath{^{\textnormal{+}}}.\vspace{0.1cm}}&\\
&&&&&\parbox[t][0.3cm]{11.07296cm}{\raggedright Decay mode: predominantly \ensuremath{^{\textnormal{14}}}Be\ensuremath{_{\textnormal{g.s.}}}+n (\href{https://www.nndc.bnl.gov/nsr/nsrlink.jsp?2013Sn02,B}{2013Sn02}, \href{https://www.nndc.bnl.gov/nsr/nsrlink.jsp?2021Co07,B}{2021Co07}, \href{https://www.nndc.bnl.gov/nsr/nsrlink.jsp?2024Ku30,B}{2024Ku30}); and\vspace{0.1cm}}&\\
&&&&&\parbox[t][0.3cm]{11.07296cm}{\raggedright {\ }{\ }{\ }a small decay branch through the \ensuremath{^{\textnormal{14}}}Be*(2\ensuremath{^{\textnormal{+}}_{\textnormal{1}}}) intermediate state (\href{https://www.nndc.bnl.gov/nsr/nsrlink.jsp?2024Ku30,B}{2024Ku30}).\vspace{0.1cm}}&\\
\multicolumn{1}{r@{}}{1700}&\multicolumn{1}{@{}l}{}&\multicolumn{1}{r@{}}{3300}&\multicolumn{1}{@{}l}{}&\multicolumn{1}{l}{\texttt{\hyperlink{BE1}{A}\ \ } }&\parbox[t][0.3cm]{11.07296cm}{\raggedright \%n\ensuremath{\approx}100 (\href{https://www.nndc.bnl.gov/nsr/nsrlink.jsp?2024Ku30,B}{2024Ku30})\vspace{0.1cm}}&\\
&&&&&\parbox[t][0.3cm]{11.07296cm}{\raggedright E\ensuremath{_{\textnormal{c.m.}}}(\ensuremath{^{\textnormal{14}}}Be\ensuremath{_{\textnormal{g.s.}}}+n) (keV): Deduced from adding S\ensuremath{_{\textnormal{2n}}}(\ensuremath{^{\textnormal{14}}}Be)=1270 keV \textit{130}\vspace{0.1cm}}&\\
&&&&&\parbox[t][0.3cm]{11.07296cm}{\raggedright {\ }{\ }{\ }(\href{https://www.nndc.bnl.gov/nsr/nsrlink.jsp?2021Wa16,B}{2021Wa16}: AME-2020) to E\ensuremath{_{\textnormal{c.m.}}}(3n+\ensuremath{^{\textnormal{12}}}Be)=2030 keV (\href{https://www.nndc.bnl.gov/nsr/nsrlink.jsp?2024Ku30,B}{2024Ku30}).\vspace{0.1cm}}&\\
&&&&&\parbox[t][0.3cm]{11.07296cm}{\raggedright E(level): This state is very weak in (\href{https://www.nndc.bnl.gov/nsr/nsrlink.jsp?2024Ku30,B}{2024Ku30}).\vspace{0.1cm}}&\\
\end{longtable}
\parbox[b][0.3cm]{17.7cm}{\makebox[1ex]{\ensuremath{^{\hypertarget{BE0LEVEL0}{a}}}} We accept E\ensuremath{_{\textnormal{c.m.}}}(3n+\ensuremath{^{\textnormal{12}}}Be)=330 keV \textit{20} (\href{https://www.nndc.bnl.gov/nsr/nsrlink.jsp?2024Ku30,B}{2024Ku30}) for \ensuremath{^{\textnormal{15}}}Be\ensuremath{_{\textnormal{g.s.}}}. This implies E\ensuremath{_{\textnormal{c.m.}}}(n+\ensuremath{^{\textnormal{14}}}Be)=1600 keV \textit{132} for \ensuremath{^{\textnormal{15}}}Be\ensuremath{_{\textnormal{g.s.}}} using}\\
\parbox[b][0.3cm]{17.7cm}{{\ }{\ }S\ensuremath{_{\textnormal{2n}}}(\ensuremath{^{\textnormal{14}}}Be)=1270 keV \textit{130} from (\href{https://www.nndc.bnl.gov/nsr/nsrlink.jsp?2021Wa16,B}{2021Wa16}: AME-2020). The energies of the excited states in \ensuremath{^{\textnormal{15}}}Be could then be deduced by}\\
\parbox[b][0.3cm]{17.7cm}{{\ }{\ }either subtracting 1600 keV \textit{132} from the relative center-of-mass energies of the associated neutrons and \ensuremath{^{\textnormal{14}}}Be decay products, or}\\
\parbox[b][0.3cm]{17.7cm}{{\ }{\ }by subtracting 330 keV \textit{20} from the relative center-of-mass energies of the associated 3n and \ensuremath{^{\textnormal{12}}}Be decay products.}\\
\vspace{0.5cm}
\clearpage
\subsection[\hspace{-0.2cm}\ensuremath{^{\textnormal{2}}}H(\ensuremath{^{\textnormal{14}}}Be,\ensuremath{^{\textnormal{15}}}Be)]{ }
\vspace{-27pt}
\vspace{0.3cm}
\hypertarget{BE1}{{\bf \small \underline{\ensuremath{^{\textnormal{2}}}H(\ensuremath{^{\textnormal{14}}}Be,\ensuremath{^{\textnormal{15}}}Be)\hspace{0.2in}\href{https://www.nndc.bnl.gov/nsr/nsrlink.jsp?2013Sn02,B}{2013Sn02},\href{https://www.nndc.bnl.gov/nsr/nsrlink.jsp?2024Ku30,B}{2024Ku30}}}}\\
\vspace{4pt}
\vspace{8pt}
\parbox[b][0.3cm]{17.7cm}{\addtolength{\parindent}{-0.2in}J\ensuremath{^{\ensuremath{\pi}}}(\ensuremath{^{\textnormal{2}}}H)=1\ensuremath{^{\textnormal{+}}} and J\ensuremath{^{\ensuremath{\pi}}}(\ensuremath{^{\textnormal{14}}}Be\ensuremath{_{\textnormal{g.s.}}})=0\ensuremath{^{\textnormal{+}}}.}\\
\parbox[b][0.3cm]{17.7cm}{\addtolength{\parindent}{-0.2in}\href{https://www.nndc.bnl.gov/nsr/nsrlink.jsp?2013Sn02,B}{2013Sn02}: The authors populated a state in the neutron unbound \ensuremath{^{\textnormal{15}}}Be nucleus and measured its decay energy. This constitutes the}\\
\parbox[b][0.3cm]{17.7cm}{first positive observation of any level in \ensuremath{^{\textnormal{15}}}Be.}\\
\parbox[b][0.3cm]{17.7cm}{\addtolength{\parindent}{-0.2in}A 59 MeV/nucleon \ensuremath{^{\textnormal{14}}}Be beam was produced by fragmentation of an \ensuremath{^{\textnormal{18}}}O beam on a \ensuremath{^{\textnormal{9}}}Be target at the National Superconducting}\\
\parbox[b][0.3cm]{17.7cm}{Cyclotron Laboratory. The \ensuremath{^{\textnormal{14}}}Be beam impinged on a 435 mg/cm\ensuremath{^{\textnormal{2}}} deuterated polyethylene target. The produced \ensuremath{^{\textnormal{15}}}Be nuclei}\\
\parbox[b][0.3cm]{17.7cm}{decayed in the target; levels decaying to the \ensuremath{^{\textnormal{14}}}Be\ensuremath{_{\textnormal{g.s.}}} were characterized by measurement of the neutron momentum (in the MoNA}\\
\parbox[b][0.3cm]{17.7cm}{array) and the \ensuremath{^{\textnormal{14}}}Be momentum (in the focal plane detectors of the sweeper magnet at \ensuremath{\theta}\ensuremath{_{\textnormal{lab}}}=43\ensuremath{^\circ}). Neutrons and \ensuremath{^{\textnormal{14}}}Be particles}\\
\parbox[b][0.3cm]{17.7cm}{were detected in coincidence mode. The kinematic reconstruction of the \ensuremath{^{\textnormal{14}}}Be+n relative energy resulted in a broad resonance at}\\
\parbox[b][0.3cm]{17.7cm}{E\ensuremath{_{\textnormal{c.m.}}}(n+\ensuremath{^{\textnormal{14}}}Be)=1.8 MeV \textit{1} with \ensuremath{\Gamma}=575 keV \textit{200}. This level was identified as the lowest J\ensuremath{^{\ensuremath{\pi}}}=5/2\ensuremath{^{\textnormal{+}}} state of \ensuremath{^{\textnormal{15}}}Be.}\\
\parbox[b][0.3cm]{17.7cm}{\addtolength{\parindent}{-0.2in}Shell model calculations predicted two states in the low-energy region of \ensuremath{^{\textnormal{15}}}Be; one with J\ensuremath{^{\ensuremath{\pi}}}=3/2\ensuremath{^{\textnormal{+}}} and another with J\ensuremath{^{\ensuremath{\pi}}}=5/2\ensuremath{^{\textnormal{+}}}. The}\\
\parbox[b][0.3cm]{17.7cm}{J\ensuremath{^{\ensuremath{\pi}}}=3/2\ensuremath{^{\textnormal{+}}} state was estimated to be unbound by at least 1.54 MeV (\href{https://www.nndc.bnl.gov/nsr/nsrlink.jsp?2011Sp01,B}{2011Sp01}) and was expected to decay to the J\ensuremath{^{\ensuremath{\pi}}}=2\ensuremath{^{\textnormal{+}}} first excited}\\
\parbox[b][0.3cm]{17.7cm}{state of \ensuremath{^{\textnormal{14}}}Be, which, in turn, decays via \ensuremath{^{\textnormal{14}}}Be*\ensuremath{\rightarrow}\ensuremath{^{\textnormal{13}}}Be+n\ensuremath{\rightarrow}\ensuremath{^{\textnormal{12}}}Be+2n. (\href{https://www.nndc.bnl.gov/nsr/nsrlink.jsp?2013Sn02,B}{2013Sn02}) acknowledged that the observation of the J\ensuremath{^{\ensuremath{\pi}}}=3/2\ensuremath{^{\textnormal{+}}}}\\
\parbox[b][0.3cm]{17.7cm}{state will be difficult.}\\
\parbox[b][0.3cm]{17.7cm}{\addtolength{\parindent}{-0.2in}The predicted order of the J\ensuremath{^{\ensuremath{\pi}}}=3/2\ensuremath{^{\textnormal{+}}} and 5/2\ensuremath{^{\textnormal{+}}} states is controversial. The present J\ensuremath{^{\ensuremath{\pi}}}=5/2\ensuremath{^{\textnormal{+}}} state was accepted as the ground state}\\
\parbox[b][0.3cm]{17.7cm}{since it was the only level observed experimentally.}\\
\parbox[b][0.3cm]{17.7cm}{\addtolength{\parindent}{-0.2in}\href{https://www.nndc.bnl.gov/nsr/nsrlink.jsp?2024Ku30,B}{2024Ku30}: In light of the theoretical work by Fortune (\href{https://www.nndc.bnl.gov/nsr/nsrlink.jsp?2015Fo04,B}{2015Fo04}, \href{https://www.nndc.bnl.gov/nsr/nsrlink.jsp?2018Fo22,B}{2018Fo22}) and to identify the predicted 3/2\ensuremath{^{\textnormal{+}}} state, (\href{https://www.nndc.bnl.gov/nsr/nsrlink.jsp?2024Ku30,B}{2024Ku30})}\\
\parbox[b][0.3cm]{17.7cm}{reanalyzed the data of (\href{https://www.nndc.bnl.gov/nsr/nsrlink.jsp?2013Sn02,B}{2013Sn02}) while emphasizing on the reconstruction of alternative decay paths such as \ensuremath{^{\textnormal{12}}}Be+n, \ensuremath{^{\textnormal{12}}}Be+2n,}\\
\parbox[b][0.3cm]{17.7cm}{and \ensuremath{^{\textnormal{12}}}Be+3n. Causality cuts with neutron multiplicity=1, 2 and 3 were applied to the data to maximize the true 2n and 3n decay}\\
\parbox[b][0.3cm]{17.7cm}{events. These cuts revealed a well-defined peak under 1 MeV in the \ensuremath{^{\textnormal{12}}}Be+3n data. A GEANT4 Monte Carlo simulation was}\\
\parbox[b][0.3cm]{17.7cm}{performed, which best described the data as the sum of the \ensuremath{^{\textnormal{14}}}Be*(2\ensuremath{^{\textnormal{+}}_{\textnormal{1}}}) state and three \ensuremath{^{\textnormal{15}}}Be states at 330, 540, and 2030 keV}\\
\parbox[b][0.3cm]{17.7cm}{above the \ensuremath{^{\textnormal{12}}}Be\ensuremath{_{\textnormal{g.s.}}}+3n threshold, respectively. The authors determined \ensuremath{\Gamma}\ensuremath{<}200 keV for the newly found resonance at}\\
\parbox[b][0.3cm]{17.7cm}{E\ensuremath{_{\textnormal{c.m.}}}(3n+\ensuremath{^{\textnormal{12}}}Be)=330 keV \textit{20}. Using the shell model predictions of (\href{https://www.nndc.bnl.gov/nsr/nsrlink.jsp?2011Sp01,B}{2011Sp01}), (\href{https://www.nndc.bnl.gov/nsr/nsrlink.jsp?2024Ku30,B}{2024Ku30}) speculated a J\ensuremath{^{\ensuremath{\pi}}}=3/2\ensuremath{^{\textnormal{+}}} assignment for}\\
\parbox[b][0.3cm]{17.7cm}{this new resonance and considered it as a candidate for the ground state of \ensuremath{^{\textnormal{15}}}Be. (\href{https://www.nndc.bnl.gov/nsr/nsrlink.jsp?2024Ku30,B}{2024Ku30}) reported that the strength of this state}\\
\parbox[b][0.3cm]{17.7cm}{relative to that of the \ensuremath{^{\textnormal{14}}}Be*(2\ensuremath{^{\textnormal{+}}_{\textnormal{1}}}) level depends on the width of this state. Assuming that this state has an optimal \ensuremath{\Gamma}=110 keV, the}\\
\parbox[b][0.3cm]{17.7cm}{strength of this state is 2.7 times stronger than the direct population of the \ensuremath{^{\textnormal{14}}}Be*(2\ensuremath{^{\textnormal{+}}_{\textnormal{1}}}) state. On the other hand, assuming that the}\\
\parbox[b][0.3cm]{17.7cm}{\ensuremath{^{\textnormal{15}}}Be\ensuremath{_{\textnormal{g.s.}}} is a narrow resonance with \ensuremath{\Gamma}\ensuremath{<}10 keV, its strength is nearly as strong as the population of \ensuremath{^{\textnormal{14}}}Be*(2\ensuremath{^{\textnormal{+}}_{\textnormal{1}}}).}\\
\parbox[b][0.3cm]{17.7cm}{\addtolength{\parindent}{-0.2in}(\href{https://www.nndc.bnl.gov/nsr/nsrlink.jsp?2013Sn02,B}{2013Sn02}) observed the J\ensuremath{^{\ensuremath{\pi}}}=5/2\ensuremath{^{\textnormal{+}}_{\textnormal{1}}} state through analysis of the \ensuremath{^{\textnormal{14}}}Be\ensuremath{_{\textnormal{g.s.}}}+n events. Based on the spectroscopic factors deduced}\\
\parbox[b][0.3cm]{17.7cm}{theoretically by (\href{https://www.nndc.bnl.gov/nsr/nsrlink.jsp?2013Sn02,B}{2013Sn02}), they suggested that the 3/2\ensuremath{^{\textnormal{+}}} state in \ensuremath{^{\textnormal{15}}}Be predominantly decays to the \ensuremath{^{\textnormal{14}}}Be*(2\ensuremath{^{\textnormal{+}}_{\textnormal{1}}}) level, while the}\\
\parbox[b][0.3cm]{17.7cm}{5/2\ensuremath{^{\textnormal{+}}_{\textnormal{1}}} state decays mainly to \ensuremath{^{\textnormal{14}}}Be\ensuremath{_{\textnormal{g.s.}}}+n. (\href{https://www.nndc.bnl.gov/nsr/nsrlink.jsp?2015Fo04,B}{2015Fo04}) disputed those spectroscopic factors and suggested that the 5/2\ensuremath{^{\textnormal{+}}_{\textnormal{1}}} state in}\\
\parbox[b][0.3cm]{17.7cm}{\ensuremath{^{\textnormal{15}}}Be would have an additional strong \textit{l}=0 decay branch to the \ensuremath{^{\textnormal{14}}}Be*(2\ensuremath{^{\textnormal{+}}_{\textnormal{1}}}) state. (\href{https://www.nndc.bnl.gov/nsr/nsrlink.jsp?2024Ku30,B}{2024Ku30}) confirmed that the decay of the 5/2\ensuremath{^{\textnormal{+}}}}\\
\parbox[b][0.3cm]{17.7cm}{state is mainly through \ensuremath{^{\textnormal{14}}}Be\ensuremath{_{\textnormal{g.s.}}}+n, but they also found a small decay branch through the \ensuremath{^{\textnormal{14}}}Be*(2\ensuremath{^{\textnormal{+}}_{\textnormal{1}}}) intermediate state.}\\
\vspace{0.385cm}
\parbox[b][0.3cm]{17.7cm}{\addtolength{\parindent}{-0.2in}\textit{Theory}:}\\
\parbox[b][0.3cm]{17.7cm}{\addtolength{\parindent}{-0.2in}\href{https://www.nndc.bnl.gov/nsr/nsrlink.jsp?2015Fo04,B}{2015Fo04}, \href{https://www.nndc.bnl.gov/nsr/nsrlink.jsp?2018Fo07,B}{2018Fo07}: Performed shell model analysis of \ensuremath{^{\textnormal{15}}}Be with an emphasis on evaluating the \textit{s}- and \textit{d}-shell single-particle}\\
\parbox[b][0.3cm]{17.7cm}{energies. The results indicated that the measured width of the proposed 5/2\ensuremath{^{\textnormal{+}}} state by (\href{https://www.nndc.bnl.gov/nsr/nsrlink.jsp?2013Sn02,B}{2013Sn02}) is 2\ensuremath{\sigma} larger than the expected}\\
\parbox[b][0.3cm]{17.7cm}{width of \ensuremath{\Gamma}=178 keV, which was calculated by (\href{https://www.nndc.bnl.gov/nsr/nsrlink.jsp?2015Fo04,B}{2015Fo04}) assuming a spectroscopic factor of S=0.44 from (\href{https://www.nndc.bnl.gov/nsr/nsrlink.jsp?2013Sn02,B}{2013Sn02}).}\\
\parbox[b][0.3cm]{17.7cm}{(\href{https://www.nndc.bnl.gov/nsr/nsrlink.jsp?2015Fo04,B}{2015Fo04}) calculated S=0.9, which resulted in \ensuremath{\Gamma}=390 keV for the \ensuremath{^{\textnormal{15}}}Be*(5/2\ensuremath{^{\textnormal{+}}}) state. They attributed the larger experimental}\\
\parbox[b][0.3cm]{17.7cm}{width to the decay of the \ensuremath{^{\textnormal{15}}}Be*(5/2\ensuremath{^{\textnormal{+}}}) state to the \ensuremath{^{\textnormal{14}}}Be*(2\ensuremath{^{\textnormal{+}}_{\textnormal{1}}}) level, whose configuration was predicted by (\href{https://www.nndc.bnl.gov/nsr/nsrlink.jsp?2015Fo04,B}{2015Fo04}) to be of \textit{ds}}\\
\parbox[b][0.3cm]{17.7cm}{nature. (\href{https://www.nndc.bnl.gov/nsr/nsrlink.jsp?2015Fo04,B}{2015Fo04}) predicted that the strong \textit{l}=0 transition from the decay of \ensuremath{^{\textnormal{15}}}Be*(5/2\ensuremath{^{\textnormal{+}}})\ensuremath{\rightarrow}\ensuremath{^{\textnormal{14}}}Be*(2\ensuremath{^{\textnormal{+}}_{\textnormal{1}}}) would be observable if the}\\
\parbox[b][0.3cm]{17.7cm}{\ensuremath{^{\textnormal{12}}}Be+3n final products from that decay are measured in coincidence.}\\
\vspace{12pt}
\underline{$^{15}$Be Levels}\\
\begin{longtable}{cccccccc@{\extracolsep{\fill}}c}
\multicolumn{2}{c}{E(level)$^{{\hyperlink{BE1LEVEL0}{a}}}$}&J$^{\pi}$$^{}$&\multicolumn{2}{c}{\ensuremath{\Gamma}$^{}$}&\multicolumn{2}{c}{E\ensuremath{_{\textnormal{c.m.}}}(\ensuremath{^{\textnormal{12}}}Be\ensuremath{_{\textnormal{g.s.}}}+3n) (keV)$^{{\hyperlink{BE1LEVEL1}{b}}}$}&Comments&\\[-.2cm]
\multicolumn{2}{c}{\hrulefill}&\hrulefill&\multicolumn{2}{c}{\hrulefill}&\multicolumn{2}{c}{\hrulefill}&\hrulefill&
\endfirsthead
\multicolumn{1}{r@{}}{0}&\multicolumn{1}{@{}l}{}&\multicolumn{1}{l}{(3/2\ensuremath{^{+}})}&\multicolumn{1}{r@{}}{$<$200}&\multicolumn{1}{@{}l}{}&\multicolumn{1}{r@{}}{330}&\multicolumn{1}{@{ }l}{{\it 20}}&\parbox[t][0.3cm]{9.62559cm}{\raggedright \%n\ensuremath{\approx}100 (\href{https://www.nndc.bnl.gov/nsr/nsrlink.jsp?2024Ku30,B}{2024Ku30})\vspace{0.1cm}}&\\
&&&&&&&\parbox[t][0.3cm]{9.62559cm}{\raggedright \ensuremath{\Gamma}: From (\href{https://www.nndc.bnl.gov/nsr/nsrlink.jsp?2024Ku30,B}{2024Ku30}): The best fit resulted in an optimal \ensuremath{\Gamma}=110 keV.\vspace{0.1cm}}&\\
&&&&&&&\parbox[t][0.3cm]{9.62559cm}{\raggedright J\ensuremath{^{\pi}}: From (\href{https://www.nndc.bnl.gov/nsr/nsrlink.jsp?2024Ku30,B}{2024Ku30}) and based on the shell model predictions by\vspace{0.1cm}}&\\
&&&&&&&\parbox[t][0.3cm]{9.62559cm}{\raggedright {\ }{\ }{\ }(\href{https://www.nndc.bnl.gov/nsr/nsrlink.jsp?2011Sp01,B}{2011Sp01}: \ensuremath{^{\textnormal{9}}}Be(\ensuremath{^{\textnormal{17}}}C,\ensuremath{^{\textnormal{15}}}Be)) and supported by the theoretical\vspace{0.1cm}}&\\
&&&&&&&\parbox[t][0.3cm]{9.62559cm}{\raggedright {\ }{\ }{\ }findings of (\href{https://www.nndc.bnl.gov/nsr/nsrlink.jsp?2015Fo04,B}{2015Fo04}, \href{https://www.nndc.bnl.gov/nsr/nsrlink.jsp?2018Fo22,B}{2018Fo22}) regarding the decay pattern of the\vspace{0.1cm}}&\\
&&&&&&&\parbox[t][0.3cm]{9.62559cm}{\raggedright {\ }{\ }{\ }5/2\ensuremath{^{\textnormal{+}}_{\textnormal{1}}} state in \ensuremath{^{\textnormal{15}}}Be.\vspace{0.1cm}}&\\
&&&&&&&\parbox[t][0.3cm]{9.62559cm}{\raggedright Decay mode: \ensuremath{^{\textnormal{14}}}Be*(2\ensuremath{^{\textnormal{+}}_{\textnormal{1}}})+n\ensuremath{\rightarrow}\ensuremath{^{\textnormal{13}}}Be\ensuremath{_{\textnormal{g.s.}}}+2n\ensuremath{\rightarrow}\ensuremath{^{\textnormal{12}}}Be\ensuremath{_{\textnormal{g.s.}}}+3n (\href{https://www.nndc.bnl.gov/nsr/nsrlink.jsp?2024Ku30,B}{2024Ku30}).\vspace{0.1cm}}&\\
&&&&&&&\parbox[t][0.3cm]{9.62559cm}{\raggedright As indicated in (\href{https://www.nndc.bnl.gov/nsr/nsrlink.jsp?2024Ku30,B}{2024Ku30}), the large \ensuremath{\Delta}E in \ensuremath{^{\textnormal{15}}}Be\ensuremath{_{\textnormal{g.s.}}} is connected to\vspace{0.1cm}}&\\
\end{longtable}
\begin{textblock}{29}(0,27.3)
Continued on next page (footnotes at end of table)
\end{textblock}
\clearpage
\begin{longtable}{cccccccc@{\extracolsep{\fill}}c}
\\[-.4cm]
\multicolumn{9}{c}{{\bf \small \underline{\ensuremath{^{\textnormal{2}}}H(\ensuremath{^{\textnormal{14}}}Be,\ensuremath{^{\textnormal{15}}}Be)\hspace{0.2in}\href{https://www.nndc.bnl.gov/nsr/nsrlink.jsp?2013Sn02,B}{2013Sn02},\href{https://www.nndc.bnl.gov/nsr/nsrlink.jsp?2024Ku30,B}{2024Ku30} (continued)}}}\\
\multicolumn{9}{c}{~}\\
\multicolumn{9}{c}{\underline{\ensuremath{^{15}}Be Levels (continued)}}\\
\multicolumn{9}{c}{~}\\
\multicolumn{2}{c}{E(level)$^{{\hyperlink{BE1LEVEL0}{a}}}$}&J$^{\pi}$$^{}$&\multicolumn{2}{c}{\ensuremath{\Gamma}$^{}$}&\multicolumn{2}{c}{E\ensuremath{_{\textnormal{c.m.}}}(\ensuremath{^{\textnormal{12}}}Be\ensuremath{_{\textnormal{g.s.}}}+3n) (keV)$^{{\hyperlink{BE1LEVEL1}{b}}}$}&Comments&\\[-.2cm]
\multicolumn{2}{c}{\hrulefill}&\hrulefill&\multicolumn{2}{c}{\hrulefill}&\multicolumn{2}{c}{\hrulefill}&\hrulefill&
\endhead
&&&&&&&\parbox[t][0.3cm]{7.9573cm}{\raggedright {\ }{\ }{\ }the \ensuremath{\Delta}S\ensuremath{_{\textnormal{2n}}}(\ensuremath{^{\textnormal{14}}}Be)=130 keV. An improvement in the\vspace{0.1cm}}&\\
&&&&&&&\parbox[t][0.3cm]{7.9573cm}{\raggedright {\ }{\ }{\ }\ensuremath{^{\textnormal{14}}}Be mass would resolve some level-energy\vspace{0.1cm}}&\\
&&&&&&&\parbox[t][0.3cm]{7.9573cm}{\raggedright {\ }{\ }{\ }uncertainties for the levels decaying via \ensuremath{^{\textnormal{14}}}Be\ensuremath{_{\textnormal{g.s.}}}+n and\vspace{0.1cm}}&\\
&&&&&&&\parbox[t][0.3cm]{7.9573cm}{\raggedright {\ }{\ }{\ }those decaying via \ensuremath{^{\textnormal{14}}}Be*(2\ensuremath{^{\textnormal{+}}_{\textnormal{1}}})+n\ensuremath{\rightarrow}\ensuremath{^{\textnormal{12}}}Be\ensuremath{_{\textnormal{g.s.}}}+3n.\vspace{0.1cm}}&\\
\multicolumn{1}{r@{}}{0.2\ensuremath{\times10^{3}}}&\multicolumn{1}{@{ }l}{{\it 2}}&\multicolumn{1}{l}{(5/2\ensuremath{^{+}})}&\multicolumn{1}{r@{}}{0}&\multicolumn{1}{@{.}l}{58 MeV {\it 20}}&\multicolumn{1}{r@{}}{540}&\multicolumn{1}{@{}l}{}&\parbox[t][0.3cm]{7.9573cm}{\raggedright E(level): From subtracting 1.60 MeV \textit{13} from\vspace{0.1cm}}&\\
&&&&&&&\parbox[t][0.3cm]{7.9573cm}{\raggedright {\ }{\ }{\ }E\ensuremath{_{\textnormal{c.m.}}}(n+\ensuremath{^{\textnormal{14}}}Be)=1.8 MeV \textit{1} (\href{https://www.nndc.bnl.gov/nsr/nsrlink.jsp?2013Sn02,B}{2013Sn02}).\vspace{0.1cm}}&\\
&&&&&&&\parbox[t][0.3cm]{7.9573cm}{\raggedright E(level): See also E\ensuremath{_{\textnormal{x}}}=210 keV, which is deduced from\vspace{0.1cm}}&\\
&&&&&&&\parbox[t][0.3cm]{7.9573cm}{\raggedright {\ }{\ }{\ }the measured value of E\ensuremath{_{\textnormal{c.m.}}}(3n+\ensuremath{^{\textnormal{12}}}Be)=540 keV\vspace{0.1cm}}&\\
&&&&&&&\parbox[t][0.3cm]{7.9573cm}{\raggedright {\ }{\ }{\ }(\href{https://www.nndc.bnl.gov/nsr/nsrlink.jsp?2024Ku30,B}{2024Ku30}). The latter value is equivalent to\vspace{0.1cm}}&\\
&&&&&&&\parbox[t][0.3cm]{7.9573cm}{\raggedright {\ }{\ }{\ }E\ensuremath{_{\textnormal{c.m.}}}(n+\ensuremath{^{\textnormal{14}}}Be)=1810 keV.\vspace{0.1cm}}&\\
&&&&&&&\parbox[t][0.3cm]{7.9573cm}{\raggedright \ensuremath{\Gamma}: From \ensuremath{\Gamma}=575 keV \textit{200} (\href{https://www.nndc.bnl.gov/nsr/nsrlink.jsp?2013Sn02,B}{2013Sn02}).\vspace{0.1cm}}&\\
&&&&&&&\parbox[t][0.3cm]{7.9573cm}{\raggedright J\ensuremath{^{\pi}}: From (\href{https://www.nndc.bnl.gov/nsr/nsrlink.jsp?2013Sn02,B}{2013Sn02}): based on the decay patterns\vspace{0.1cm}}&\\
&&&&&&&\parbox[t][0.3cm]{7.9573cm}{\raggedright {\ }{\ }{\ }predicted by their shell-model calculations and the\vspace{0.1cm}}&\\
&&&&&&&\parbox[t][0.3cm]{7.9573cm}{\raggedright {\ }{\ }{\ }spectroscopic overlap of the proposed \ensuremath{^{\textnormal{15}}}Be levels with\vspace{0.1cm}}&\\
&&&&&&&\parbox[t][0.3cm]{7.9573cm}{\raggedright {\ }{\ }{\ }\ensuremath{^{\textnormal{14}}}Be\ensuremath{_{\textnormal{g.s.}}}, which were calculated by (\href{https://www.nndc.bnl.gov/nsr/nsrlink.jsp?2013Sn02,B}{2013Sn02}) using\vspace{0.1cm}}&\\
&&&&&&&\parbox[t][0.3cm]{7.9573cm}{\raggedright {\ }{\ }{\ }FRESCO: S=0.44 if J\ensuremath{^{\ensuremath{\pi}}}=5/2\ensuremath{^{\textnormal{+}}} and S=0.07 if J\ensuremath{^{\ensuremath{\pi}}}=3/2\ensuremath{^{\textnormal{+}}}.\vspace{0.1cm}}&\\
\multicolumn{1}{r@{}}{1700}&\multicolumn{1}{@{}l}{}&&&&\multicolumn{1}{r@{}}{2030}&\multicolumn{1}{@{}l}{}&\parbox[t][0.3cm]{7.9573cm}{\raggedright \%n\ensuremath{\approx}100 (\href{https://www.nndc.bnl.gov/nsr/nsrlink.jsp?2024Ku30,B}{2024Ku30})\vspace{0.1cm}}&\\
&&&&&&&\parbox[t][0.3cm]{7.9573cm}{\raggedright E(level): From subtracting 330 keV \textit{20} from\vspace{0.1cm}}&\\
&&&&&&&\parbox[t][0.3cm]{7.9573cm}{\raggedright {\ }{\ }{\ }E\ensuremath{_{\textnormal{c.m.}}}(3n+\ensuremath{^{\textnormal{12}}}Be)=2030 keV (\href{https://www.nndc.bnl.gov/nsr/nsrlink.jsp?2024Ku30,B}{2024Ku30}).\vspace{0.1cm}}&\\
&&&&&&&\parbox[t][0.3cm]{7.9573cm}{\raggedright This state is very weak in (\href{https://www.nndc.bnl.gov/nsr/nsrlink.jsp?2024Ku30,B}{2024Ku30}).\vspace{0.1cm}}&\\
\end{longtable}
\parbox[b][0.3cm]{17.7cm}{\makebox[1ex]{\ensuremath{^{\hypertarget{BE1LEVEL0}{a}}}} We accept E\ensuremath{_{\textnormal{c.m.}}}(3n+\ensuremath{^{\textnormal{12}}}Be)=330 keV \textit{20} (\href{https://www.nndc.bnl.gov/nsr/nsrlink.jsp?2024Ku30,B}{2024Ku30}) for \ensuremath{^{\textnormal{15}}}Be\ensuremath{_{\textnormal{g.s.}}}. This implies E\ensuremath{_{\textnormal{c.m.}}}(n+\ensuremath{^{\textnormal{14}}}Be)=1600 keV \textit{132} for \ensuremath{^{\textnormal{15}}}Be\ensuremath{_{\textnormal{g.s.}}} using}\\
\parbox[b][0.3cm]{17.7cm}{{\ }{\ }S\ensuremath{_{\textnormal{2n}}}(\ensuremath{^{\textnormal{14}}}Be)=1270 keV \textit{130} from (\href{https://www.nndc.bnl.gov/nsr/nsrlink.jsp?2021Wa16,B}{2021Wa16}: AME-2020). The energies of the excited states in \ensuremath{^{\textnormal{15}}}Be could then be deduced by}\\
\parbox[b][0.3cm]{17.7cm}{{\ }{\ }either subtracting 1600 keV \textit{132} from the relative center-of-mass energies of the associated neutrons and \ensuremath{^{\textnormal{14}}}Be decay products, or}\\
\parbox[b][0.3cm]{17.7cm}{{\ }{\ }by subtracting 330 keV \textit{20} from the center-of-mass relative energies of 3n and \ensuremath{^{\textnormal{12}}}Be decay products.}\\
\parbox[b][0.3cm]{17.7cm}{\makebox[1ex]{\ensuremath{^{\hypertarget{BE1LEVEL1}{b}}}} From (\href{https://www.nndc.bnl.gov/nsr/nsrlink.jsp?2024Ku30,B}{2024Ku30}).}\\
\vspace{0.5cm}
\clearpage
\subsection[\hspace{-0.2cm}\ensuremath{^{\textnormal{9}}}Be(\ensuremath{^{\textnormal{17}}}C,\ensuremath{^{\textnormal{15}}}Be2p)]{ }
\vspace{-27pt}
\vspace{0.3cm}
\hypertarget{BE2}{{\bf \small \underline{\ensuremath{^{\textnormal{9}}}Be(\ensuremath{^{\textnormal{17}}}C,\ensuremath{^{\textnormal{15}}}Be2p)\hspace{0.2in}\href{https://www.nndc.bnl.gov/nsr/nsrlink.jsp?2011Sp01,B}{2011Sp01},\href{https://www.nndc.bnl.gov/nsr/nsrlink.jsp?2015Ku04,B}{2015Ku04}}}}\\
\vspace{4pt}
\vspace{8pt}
\parbox[b][0.3cm]{17.7cm}{\addtolength{\parindent}{-0.2in}Two proton knockout reaction.}\\
\parbox[b][0.3cm]{17.7cm}{\addtolength{\parindent}{-0.2in}J\ensuremath{^{\ensuremath{\pi}}}(\ensuremath{^{\textnormal{9}}}Be\ensuremath{_{\textnormal{g.s.}}})=3/2\ensuremath{^{-}} and J\ensuremath{^{\ensuremath{\pi}}}(\ensuremath{^{\textnormal{17}}}C)=3/2\ensuremath{^{\textnormal{+}}}.}\\
\parbox[b][0.3cm]{17.7cm}{\addtolength{\parindent}{-0.2in}\href{https://www.nndc.bnl.gov/nsr/nsrlink.jsp?2008SpZV,B}{2008SpZV}, \href{https://www.nndc.bnl.gov/nsr/nsrlink.jsp?2011Sp01,B}{2011Sp01}: This work was motivated by a study of the \ensuremath{^{\textnormal{16}}}Be\ensuremath{_{\textnormal{g.s.}}} decay mechanism, which could be expected to 1-n or}\\
\parbox[b][0.3cm]{17.7cm}{2-n decay, depending on the \ensuremath{^{\textnormal{15}}}Be mass.}\\
\parbox[b][0.3cm]{17.7cm}{\addtolength{\parindent}{-0.2in}A beam of 55 MeV/A \ensuremath{^{\textnormal{17}}}C ions impinged on a 470 mg/cm\ensuremath{^{\textnormal{2}}} \ensuremath{^{\textnormal{9}}}Be target at the NSCL MoNA/Sweeper dipole magnet target position.}\\
\parbox[b][0.3cm]{17.7cm}{Following 2p removal events in the \ensuremath{^{\textnormal{9}}}Be target, the experiment was configured to measure the momenta of \ensuremath{^{\textnormal{14}}}Be ions using the}\\
\parbox[b][0.3cm]{17.7cm}{sweeper dipole magnet and the momenta of neutrons using the MoNA neutron array. No peaks were observed in the kinematic}\\
\parbox[b][0.3cm]{17.7cm}{reconstruction of \ensuremath{^{\textnormal{14}}}Be + neutron events. The authors discuss the possible case where \ensuremath{^{\textnormal{15}}}Be decays to the \ensuremath{^{\textnormal{14}}}Be*(1.54 MeV) state,}\\
\parbox[b][0.3cm]{17.7cm}{which is known to decay to \ensuremath{^{\textnormal{12}}}Be+2n. However, the statistics were not sufficient to analyze the \ensuremath{^{\textnormal{12}}}Be+3n events. It was suggested}\\
\parbox[b][0.3cm]{17.7cm}{that \ensuremath{^{\textnormal{15}}}Be must be unbound by 1.54 MeV for this decay to occur.}\\
\parbox[b][0.3cm]{17.7cm}{\addtolength{\parindent}{-0.2in}\href{https://www.nndc.bnl.gov/nsr/nsrlink.jsp?2015Ku04,B}{2015Ku04}: Fragmentation of a \ensuremath{^{\textnormal{22}}}Ne beam at the NSCL produced a 55 MeV/nucleon \ensuremath{^{\textnormal{17}}}C beam, which impinged on a \ensuremath{^{\textnormal{9}}}Be target.}\\
\parbox[b][0.3cm]{17.7cm}{The \ensuremath{^{\textnormal{12}}}Be+neutrons decay products were measured in coincidence using the Sweeper dipole magnet and MoNA neutron array. The}\\
\parbox[b][0.3cm]{17.7cm}{relative energy spectra of the \ensuremath{^{\textnormal{12}}}Be+neutrons with neutron multiplicity of up to 3 were reconstructed. The contributions from}\\
\parbox[b][0.3cm]{17.7cm}{multiple neutron scattering was diminished using causality cuts to the data with neutron multiplicity=1 and 2. However, low}\\
\parbox[b][0.3cm]{17.7cm}{statistics prevented such a cut on neutron multiplicity=3 events. The \ensuremath{^{\textnormal{12}}}Be+2n data show evidence for the population of the}\\
\parbox[b][0.3cm]{17.7cm}{\ensuremath{^{\textnormal{14}}}Be*(1.54 MeV, 2\ensuremath{^{\textnormal{+}}}) level, but no convincing evidence was found for any \ensuremath{^{\textnormal{15}}}Be state. In the data, there is a weak suggestion of a}\\
\parbox[b][0.3cm]{17.7cm}{state at E\ensuremath{_{\textnormal{c.m.}}}(3n+\ensuremath{^{\textnormal{12}}}Be)\ensuremath{\approx}1.4 MeV that decays via the \ensuremath{^{\textnormal{14}}}Be*(1.54 MeV, 2\ensuremath{^{\textnormal{+}}}) level; however the data are well fitted without}\\
\parbox[b][0.3cm]{17.7cm}{including this state.}\\
\vspace{12pt}
\underline{$^{15}$Be Levels}\\
\vspace{0.34cm}
\parbox[b][0.3cm]{17.7cm}{\addtolength{\parindent}{-0.254cm}(\href{https://www.nndc.bnl.gov/nsr/nsrlink.jsp?2011Sp01,B}{2011Sp01}) reported that the ground state of \ensuremath{^{\textnormal{15}}}Be must be at E\ensuremath{_{\textnormal{c.m.}}}(n+\ensuremath{^{\textnormal{14}}}Be)\ensuremath{>}1.54 MeV.}\\
\vspace{0.34cm}
\begin{longtable}{cccccc@{\extracolsep{\fill}}c}
\multicolumn{2}{c}{E(level)$^{}$}&J$^{\pi}$$^{}$&\multicolumn{2}{c}{E\ensuremath{_{\textnormal{c.m.}}}(\ensuremath{^{\textnormal{12}}}Be\ensuremath{_{\textnormal{g.s.}}}+3n) (MeV)$^{{\hyperlink{BE2LEVEL0}{a}}}$}&Comments&\\[-.2cm]
\multicolumn{2}{c}{\hrulefill}&\hrulefill&\multicolumn{2}{c}{\hrulefill}&\hrulefill&
\endfirsthead
\multicolumn{1}{r@{}}{1.1\ensuremath{\times10^{3}}?}&\multicolumn{1}{@{}l}{}&\multicolumn{1}{l}{(3/2\ensuremath{^{+}})}&\multicolumn{1}{r@{}}{1}&\multicolumn{1}{@{.}l}{4}&\parbox[t][0.3cm]{10.76694cm}{\raggedright E(level): No evidence is reported in (\href{https://www.nndc.bnl.gov/nsr/nsrlink.jsp?2024Ku30,B}{2024Ku30}) for this tentative level.\vspace{0.1cm}}&\\
&&&&&\parbox[t][0.3cm]{10.76694cm}{\raggedright {\ }{\ }{\ }Therefore, this tentative state is not considered in the \ensuremath{^{\textnormal{15}}}Be Adopted Levels.\vspace{0.1cm}}&\\
&&&&&\parbox[t][0.3cm]{10.76694cm}{\raggedright E(level): The data of (\href{https://www.nndc.bnl.gov/nsr/nsrlink.jsp?2015Ku04,B}{2015Ku04}) can be fitted with the inclusion of a state at\vspace{0.1cm}}&\\
&&&&&\parbox[t][0.3cm]{10.76694cm}{\raggedright {\ }{\ }{\ }E\ensuremath{_{\textnormal{c.m.}}}(\ensuremath{^{\textnormal{12}}}Be+3n)\ensuremath{\approx}1.4 MeV or E\ensuremath{_{\textnormal{c.m.}}}(n+\ensuremath{^{\textnormal{14}}}Be)=2.66 MeV (\href{https://www.nndc.bnl.gov/nsr/nsrlink.jsp?2015Ku04,B}{2015Ku04}). This\vspace{0.1cm}}&\\
&&&&&\parbox[t][0.3cm]{10.76694cm}{\raggedright {\ }{\ }{\ }tentative excited state, which is not firmly accepted by the authors, would be\vspace{0.1cm}}&\\
&&&&&\parbox[t][0.3cm]{10.76694cm}{\raggedright {\ }{\ }{\ }at E\ensuremath{_{\textnormal{x}}}=1.1 MeV deduced from subtracting E\ensuremath{_{\textnormal{c.m.}}}(3n+\ensuremath{^{\textnormal{12}}}Be)=330 keV \textit{20}\vspace{0.1cm}}&\\
&&&&&\parbox[t][0.3cm]{10.76694cm}{\raggedright {\ }{\ }{\ }(\href{https://www.nndc.bnl.gov/nsr/nsrlink.jsp?2024Ku30,B}{2024Ku30}) from E\ensuremath{_{\textnormal{rel}}}(3n+\ensuremath{^{\textnormal{12}}}Be)=1.4 MeV from (\href{https://www.nndc.bnl.gov/nsr/nsrlink.jsp?2015Ku04,B}{2015Ku04}).\vspace{0.1cm}}&\\
&&&&&\parbox[t][0.3cm]{10.76694cm}{\raggedright J\ensuremath{^{\pi}}: From speculations in (\href{https://www.nndc.bnl.gov/nsr/nsrlink.jsp?2015Ku04,B}{2015Ku04}).\vspace{0.1cm}}&\\
&&&&&\parbox[t][0.3cm]{10.76694cm}{\raggedright (\href{https://www.nndc.bnl.gov/nsr/nsrlink.jsp?2015Ku04,B}{2015Ku04}) reported that this state, if existed, decays to \ensuremath{^{\textnormal{14}}}Be*(1.54 MeV,\vspace{0.1cm}}&\\
&&&&&\parbox[t][0.3cm]{10.76694cm}{\raggedright {\ }{\ }{\ }2\ensuremath{^{\textnormal{+}}})+n\ensuremath{\rightarrow}\ensuremath{^{\textnormal{12}}}Be+3n with a probability of \ensuremath{\leq}11\%.\vspace{0.1cm}}&\\
\end{longtable}
\parbox[b][0.3cm]{17.7cm}{\makebox[1ex]{\ensuremath{^{\hypertarget{BE2LEVEL0}{a}}}} From (\href{https://www.nndc.bnl.gov/nsr/nsrlink.jsp?2015Ku04,B}{2015Ku04}).}\\
\vspace{0.5cm}
\clearpage
\subsection[\hspace{-0.2cm}\ensuremath{^{\textnormal{12}}}C(\ensuremath{^{\textnormal{18}}}C,\ensuremath{^{\textnormal{15}}}Be)]{ }
\vspace{-27pt}
\vspace{0.3cm}
\hypertarget{BE3}{{\bf \small \underline{\ensuremath{^{\textnormal{12}}}C(\ensuremath{^{\textnormal{18}}}C,\ensuremath{^{\textnormal{15}}}Be)\hspace{0.2in}\href{https://www.nndc.bnl.gov/nsr/nsrlink.jsp?2021Co07,B}{2021Co07}}}}\\
\vspace{4pt}
\vspace{8pt}
\parbox[b][0.3cm]{17.7cm}{\addtolength{\parindent}{-0.2in}J\ensuremath{^{\ensuremath{\pi}}}(\ensuremath{^{\textnormal{12}}}C\ensuremath{_{\textnormal{g.s.}}})=0\ensuremath{^{\textnormal{+}}} and J\ensuremath{^{\ensuremath{\pi}}}(\ensuremath{^{\textnormal{18}}}C\ensuremath{_{\textnormal{g.s.}}})=0\ensuremath{^{\textnormal{+}}}.}\\
\parbox[b][0.3cm]{17.7cm}{\addtolength{\parindent}{-0.2in}A 345 MeV/nucleon \ensuremath{^{\textnormal{48}}}Ca beam was fragmented on a \ensuremath{^{\textnormal{9}}}Be target at the Radioactive Isotope Beam Factory in RIKEN. The resulting}\\
\parbox[b][0.3cm]{17.7cm}{cocktail beam included 242 MeV/nucleon \ensuremath{^{\textnormal{18}}}C and 260 MeV/nucleon \ensuremath{^{\textnormal{17}}}B ions. The experiment consisted of two phases: Day-One}\\
\parbox[b][0.3cm]{17.7cm}{and SAMURAI18 experimental campaigns.}\\
\parbox[b][0.3cm]{17.7cm}{\addtolength{\parindent}{-0.2in}During the SAMURAI18 campaign, the \ensuremath{^{\textnormal{17}}}B beam was identified by the ToF detection system of the BigRIPS fragment separator.}\\
\parbox[b][0.3cm]{17.7cm}{This selected beam impinged on the liquid hydrogen target of the MINOS Time Projection Chamber, which was surrounded by the}\\
\parbox[b][0.3cm]{17.7cm}{DALI2 NaI \ensuremath{\gamma}-detector array. The \ensuremath{^{\textnormal{14}}}Be decay products from \ensuremath{^{\textnormal{17}}}B(p,2p)\ensuremath{^{\textnormal{16}}}Be\ensuremath{\rightarrow}\ensuremath{^{\textnormal{14}}}Be+2n were momentum analyzed and measured by}\\
\parbox[b][0.3cm]{17.7cm}{the SAMURAI dipole magnet and its associated focal-plane detection system. The momenta of the associated neutrons were}\\
\parbox[b][0.3cm]{17.7cm}{measured using two walls formed by the NEBULA plastic scintillator array. The analysis of the \ensuremath{^{\textnormal{14}}}Be+2n relative{\textminus}energy spectrum}\\
\parbox[b][0.3cm]{17.7cm}{revealed that no \ensuremath{^{\textnormal{15}}}Be states were populated and that the populated \ensuremath{^{\textnormal{16}}}Be(g.s., 1.5 MeV) states seem to decay directly to \ensuremath{^{\textnormal{14}}}Be\ensuremath{_{\textnormal{g.s.}}}}\\
\parbox[b][0.3cm]{17.7cm}{(see Belen Monteagudo Godoy, Thesis (Normandie Universit\'{e}, 2019)}\\
\parbox[b][0.3cm]{17.7cm}{\textit{Structure and Neutron Decay of the Unbound Beryllium Isotopes} \ensuremath{^{\textnormal{15,16}}}\textit{Be}; and \textit{Recent Progress in Few-Body Physics}, p. 331}\\
\parbox[b][0.3cm]{17.7cm}{(2020)).}\\
\parbox[b][0.3cm]{17.7cm}{\addtolength{\parindent}{-0.2in}During the Day-One campaign, the \ensuremath{^{\textnormal{18}}}C beam was selected using the BigRIPS separator. This beam impinged on a 1.8-g/cm\ensuremath{^{\textnormal{2}}}-thick}\\
\parbox[b][0.3cm]{17.7cm}{carbon target, which was surrounded by the DALI2 NaI \ensuremath{\gamma}-detector array. The \ensuremath{^{\textnormal{15}}}Be nuclei produced from the fragmentation of the}\\
\parbox[b][0.3cm]{17.7cm}{\ensuremath{^{\textnormal{18}}}C beam decayed via \ensuremath{^{\textnormal{14}}}Be+n. These decay products were measured in coincidence using the detection system of SAMURAI and}\\
\parbox[b][0.3cm]{17.7cm}{the NEBULA array, respectively. The n+\ensuremath{^{\textnormal{14}}}Be relative energy spectrum was deduced using the invariant mass technique. The}\\
\parbox[b][0.3cm]{17.7cm}{spectrum displayed two significant peak structures at E\ensuremath{_{\textnormal{rel.}}}(n+\ensuremath{^{\textnormal{14}}}Be)\ensuremath{\approx}0.5 and 1.7 MeV. The non-resonant contribution to the}\\
\parbox[b][0.3cm]{17.7cm}{n+\ensuremath{^{\textnormal{14}}}Be pairs from the \ensuremath{^{\textnormal{16}}}Be\ensuremath{\rightarrow}2n+\ensuremath{^{\textnormal{14}}}Be events described the E\ensuremath{_{\textnormal{rel.}}}=0.5 MeV peak since \ensuremath{^{\textnormal{16}}}Be\ensuremath{_{\textnormal{g.s.}}} decays directly to \ensuremath{^{\textnormal{14}}}Be+2n with}\\
\parbox[b][0.3cm]{17.7cm}{no known sequential branch via \ensuremath{^{\textnormal{15}}}Be+n. The remaining E\ensuremath{_{\textnormal{rel.}}}=1.70 MeV \textit{13} peak has \ensuremath{\Gamma}\ensuremath{\geq}200 keV and is associated with the \ensuremath{^{\textnormal{15}}}Be}\\
\parbox[b][0.3cm]{17.7cm}{state previously reported at E\ensuremath{_{\textnormal{rel.}}}=1.8 MeV \textit{1} with J\ensuremath{^{\ensuremath{\pi}}}=(5/2\ensuremath{^{\textnormal{+}}}) by (\href{https://www.nndc.bnl.gov/nsr/nsrlink.jsp?2013Sn02,B}{2013Sn02}).}\\
\parbox[b][0.3cm]{17.7cm}{\addtolength{\parindent}{-0.2in}(\href{https://www.nndc.bnl.gov/nsr/nsrlink.jsp?2021Co07,B}{2021Co07}) suggested that the expected J\ensuremath{^{\ensuremath{\pi}}}=1/2\ensuremath{^{\textnormal{+}}} and 3/2\ensuremath{^{\textnormal{+}}} states, which remained unobserved in their work, could either be}\\
\parbox[b][0.3cm]{17.7cm}{completely degenerate with the J\ensuremath{^{\ensuremath{\pi}}}=5/2\ensuremath{^{\textnormal{+}}} level, or that they decay exclusively to \ensuremath{^{\textnormal{12}}}Be+3n and would thus be absent from the}\\
\parbox[b][0.3cm]{17.7cm}{investigated \ensuremath{^{\textnormal{14}}}Be+n channel.}\\
\vspace{12pt}
\underline{$^{15}$Be Levels}\\
\begin{longtable}{ccccccc@{\extracolsep{\fill}}c}
\multicolumn{2}{c}{E(level)$^{}$}&\multicolumn{2}{c}{\ensuremath{\Gamma}$^{}$}&\multicolumn{2}{c}{E\ensuremath{_{\textnormal{rel.}}}(n+\ensuremath{^{\textnormal{14}}}Be)(MeV)$^{}$}&Comments&\\[-.2cm]
\multicolumn{2}{c}{\hrulefill}&\multicolumn{2}{c}{\hrulefill}&\multicolumn{2}{c}{\hrulefill}&\hrulefill&
\endfirsthead
\multicolumn{1}{r@{}}{0.10\ensuremath{\times10^{3}}}&\multicolumn{1}{@{}l}{}&\multicolumn{1}{r@{}}{$\geq$200}&\multicolumn{1}{@{ }l}{keV}&\multicolumn{1}{r@{}}{1}&\multicolumn{1}{@{.}l}{70 {\it 13}}&\parbox[t][0.3cm]{10.731791cm}{\raggedright The discovery of \ensuremath{^{\textnormal{15}}}Be is reported in (\href{https://www.nndc.bnl.gov/nsr/nsrlink.jsp?2013Sn02,B}{2013Sn02}) by finding this state. It was\vspace{0.1cm}}&\\
&&&&&&\parbox[t][0.3cm]{10.731791cm}{\raggedright {\ }{\ }{\ }considered as the ground state. However, (\href{https://www.nndc.bnl.gov/nsr/nsrlink.jsp?2024Ku30,B}{2024Ku30}) found a lower-lying\vspace{0.1cm}}&\\
&&&&&&\parbox[t][0.3cm]{10.731791cm}{\raggedright {\ }{\ }{\ }state that is a candidate for the ground state making this level an excited\vspace{0.1cm}}&\\
&&&&&&\parbox[t][0.3cm]{10.731791cm}{\raggedright {\ }{\ }{\ }state.\vspace{0.1cm}}&\\
&&&&&&\parbox[t][0.3cm]{10.731791cm}{\raggedright E(level): Deduced from subtracting 1.60 MeV \textit{13} from E\ensuremath{_{\textnormal{rel}}}(n+\ensuremath{^{\textnormal{14}}}Be)=1.70\vspace{0.1cm}}&\\
&&&&&&\parbox[t][0.3cm]{10.731791cm}{\raggedright {\ }{\ }{\ }MeV \textit{13} (\href{https://www.nndc.bnl.gov/nsr/nsrlink.jsp?2021Co17,B}{2021Co17}).\vspace{0.1cm}}&\\
&&&&&&\parbox[t][0.3cm]{10.731791cm}{\raggedright \ensuremath{\Gamma},E\ensuremath{_{\textnormal{rel.}}}(n+\ensuremath{^{\textnormal{14}}}Be)(MeV): From (\href{https://www.nndc.bnl.gov/nsr/nsrlink.jsp?2021Co07,B}{2021Co07}).\vspace{0.1cm}}&\\
\end{longtable}
\end{center}
\clearpage
\newpage
\pagestyle{plain}
\section[References]{ }
\vspace{-30pt}
\begin{longtable}{l@{\hskip 0.9cm}l}
\multicolumn{2}{c}{REFERENCES FOR A=15}\\
&\endfirsthead
\multicolumn{2}{c}{REFERENCES FOR A=15(CONTINUED)}\\
&\endhead
\href{https://www.nndc.bnl.gov/nsr/nsrlink.jsp?1981Se06,B}{1981Se06}&\parbox[t]{6in}{\addtolength{\parindent}{-0.25cm}M.Seya, M.Kohno, S.Nagata - Prog.Theor.Phys.(Kyoto) 65, 204 (1981).}\\
&\parbox[t]{6in}{\addtolength{\parindent}{-0.25cm} \textit{Nuclear Binding Mechanism and Structure of Neutron-Rich Be and B Isotopes by Molecular-Orbital Model.}}\\
\href{https://www.nndc.bnl.gov/nsr/nsrlink.jsp?1985Po10,B}{1985Po10}&\parbox[t]{6in}{\addtolength{\parindent}{-0.25cm}N.A.F.M.Poppelier, L.D.Wood, P.W.M.Glaudemans - Phys.Lett. 157B, 120 (1985).}\\
&\parbox[t]{6in}{\addtolength{\parindent}{-0.25cm} \textit{Properties of Exotic p-Shell Nuclei.}}\\
\href{https://www.nndc.bnl.gov/nsr/nsrlink.jsp?1987Sa15,B}{1987Sa15}&\parbox[t]{6in}{\addtolength{\parindent}{-0.25cm}H.Sagawa, H.Toki - J.Phys.(London) G13, 453 (1987).}\\
&\parbox[t]{6in}{\addtolength{\parindent}{-0.25cm} \textit{Hartree-Fock Calculations of Light Neutron-Rich Nuclei.}}\\
\href{https://www.nndc.bnl.gov/nsr/nsrlink.jsp?2006Ko02,B}{2006Ko02}&\parbox[t]{6in}{\addtolength{\parindent}{-0.25cm}V.B.Kopeliovich, A.M.Shunderuk, G.K.Matushko - Phys.Atomic Nuclei 69, 120 (2006).}\\
&\parbox[t]{6in}{\addtolength{\parindent}{-0.25cm} \textit{Mass Splittings of Nuclear Isotopes in Chiral Soliton Approach.}}\\
\href{https://www.nndc.bnl.gov/nsr/nsrlink.jsp?2008SpZV,B}{2008SpZV}&\parbox[t]{6in}{\addtolength{\parindent}{-0.25cm}A.Spyrou, T.Baumann, D.Bazin, E.Breitbach et al. - Book of Abstracts, 12th in series of nuclear structure 2008, MichiganState University, East Lansing, Michigan, June 3-6, 2008 p.177 (2008).}\\
&\parbox[t]{6in}{\addtolength{\parindent}{-0.25cm} \textit{Measurement of the \ensuremath{^{\textnormal{15}}}Be ground state.}}\\
\href{https://www.nndc.bnl.gov/nsr/nsrlink.jsp?2011Pr03,B}{2011Pr03}&\parbox[t]{6in}{\addtolength{\parindent}{-0.25cm}B.Pritychenko, E.Betak, M.A.Kellett, B.Singh, J.Totans - Nucl.Instrum.Methods Phys.Res. A640, 213 (2011).}\\
&\parbox[t]{6in}{\addtolength{\parindent}{-0.25cm} \textit{The Nuclear Science References (NSR) database and Web Retrieval System.}}\\
\href{https://www.nndc.bnl.gov/nsr/nsrlink.jsp?2011Sp01,B}{2011Sp01}&\parbox[t]{6in}{\addtolength{\parindent}{-0.25cm}A.Spyrou, J.K.Smith, T.Baumann, B.A.Brown et al. - Phys.Rev. C 84, 044309 (2011).}\\
&\parbox[t]{6in}{\addtolength{\parindent}{-0.25cm} \textit{Search for the \ensuremath{^{\textnormal{15}}}Be ground state.}}\\
\href{https://www.nndc.bnl.gov/nsr/nsrlink.jsp?2013Sn02,B}{2013Sn02}&\parbox[t]{6in}{\addtolength{\parindent}{-0.25cm}J.Snyder, T.Baumann, G.Christian, R.A.Haring-Kaye et al. - Phys.Rev. C 88, 031303 (2013).}\\
&\parbox[t]{6in}{\addtolength{\parindent}{-0.25cm} \textit{First observation of \ensuremath{^{\textnormal{15}}}Be.}}\\
\href{https://www.nndc.bnl.gov/nsr/nsrlink.jsp?2015Fo04,B}{2015Fo04}&\parbox[t]{6in}{\addtolength{\parindent}{-0.25cm}H.T.Fortune - Phys.Rev. C 91, 034314 (2015).}\\
&\parbox[t]{6in}{\addtolength{\parindent}{-0.25cm} \textit{Properties of \ensuremath{^{\textnormal{15}}}Be (5/2\ensuremath{^{\textnormal{+}}}).}}\\
\href{https://www.nndc.bnl.gov/nsr/nsrlink.jsp?2015Ku04,B}{2015Ku04}&\parbox[t]{6in}{\addtolength{\parindent}{-0.25cm}A.N.Kuchera, A.Spyrou, J.K.Smith, T.Baumann et al. - Phys.Rev. C 91, 017304 (2015).}\\
&\parbox[t]{6in}{\addtolength{\parindent}{-0.25cm} \textit{Search for unbound \ensuremath{^{\textnormal{15}}}Be states in the 3n+ \ensuremath{^{\textnormal{12}}}Be channel.}}\\
\href{https://www.nndc.bnl.gov/nsr/nsrlink.jsp?2018Du01,B}{2018Du01}&\parbox[t]{6in}{\addtolength{\parindent}{-0.25cm}S.K.Dutta, D.Gupta, S.K.Saha - Phys.Lett. B 776, 464 (2018).}\\
&\parbox[t]{6in}{\addtolength{\parindent}{-0.25cm} \textit{Resonance state wave functions of \ensuremath{^{\textnormal{15}}}Be using supersymmetric quantummechanics.}}\\
\href{https://www.nndc.bnl.gov/nsr/nsrlink.jsp?2018Fo07,B}{2018Fo07}&\parbox[t]{6in}{\addtolength{\parindent}{-0.25cm}H.T.Fortune - Eur.Phys.J. A 54, 51 (2018).}\\
&\parbox[t]{6in}{\addtolength{\parindent}{-0.25cm} \textit{Structure of exotic light nuclei: Z = 2, 3, 4.}}\\
\href{https://www.nndc.bnl.gov/nsr/nsrlink.jsp?2018Fo22,B}{2018Fo22}&\parbox[t]{6in}{\addtolength{\parindent}{-0.25cm}H.T.Fortune - Phys.Rev. C 98, 054317 (2018).}\\
&\parbox[t]{6in}{\addtolength{\parindent}{-0.25cm} \textit{Decays of \ensuremath{^{\textnormal{15}}}Be(5/2\ensuremath{^{\textnormal{+}}}).}}\\
\href{https://www.nndc.bnl.gov/nsr/nsrlink.jsp?2019Fo09,B}{2019Fo09}&\parbox[t]{6in}{\addtolength{\parindent}{-0.25cm}H.T.Fortune - Phys.Rev. C 99, 044318 (2019).}\\
&\parbox[t]{6in}{\addtolength{\parindent}{-0.25cm} \textit{2n decays of \ensuremath{^{\textnormal{16}}}Be.}}\\
\href{https://www.nndc.bnl.gov/nsr/nsrlink.jsp?2020Wi12,B}{2020Wi12}&\parbox[t]{6in}{\addtolength{\parindent}{-0.25cm}M.Wiedeking, A.O.Macchiavelli - Eur.Phys.J. A 56, 285 (2020).}\\
&\parbox[t]{6in}{\addtolength{\parindent}{-0.25cm} \textit{Global trends of nuclear d\ensuremath{^{\textnormal{2,3,4}}_{\textnormal{5/2}}} configurations, Application of a simple effective-interaction model.}}\\
\href{https://www.nndc.bnl.gov/nsr/nsrlink.jsp?2021Co07,B}{2021Co07}&\parbox[t]{6in}{\addtolength{\parindent}{-0.25cm}A.Corsi, B.Monteagudo, F.M.Marques - Eur.Phys.J. A 57, 88 (2021).}\\
&\parbox[t]{6in}{\addtolength{\parindent}{-0.25cm} \textit{The neutron dripline at Z = 4: the case of \ensuremath{^{\textnormal{13,15}}}Be.}}\\
\href{https://www.nndc.bnl.gov/nsr/nsrlink.jsp?2021Co17,B}{2021Co17}&\parbox[t]{6in}{\addtolength{\parindent}{-0.25cm}V.D.Cong, T.D.Xuan, N.X.Hai, P.D.Khang et al. - Eur.Phys.J. A 57, 304 (2021).}\\
&\parbox[t]{6in}{\addtolength{\parindent}{-0.25cm} \textit{Normalizing the enhanced generalized superfluid model of nuclear leveldensity.}}\\
\href{https://www.nndc.bnl.gov/nsr/nsrlink.jsp?2021Wa16,B}{2021Wa16}&\parbox[t]{6in}{\addtolength{\parindent}{-0.25cm}M.Wang, W.J.Huang, F.G.Kondev, G.Audi, S.Naimi - Chin.Phys.C 45, 030003 (2021).}\\
&\parbox[t]{6in}{\addtolength{\parindent}{-0.25cm} \textit{The AME 2020 atomic mass evaluation (II). Tables, graphs and references.}}\\
\href{https://www.nndc.bnl.gov/nsr/nsrlink.jsp?2024Ku30,B}{2024Ku30}&\parbox[t]{6in}{\addtolength{\parindent}{-0.25cm}A.N.Kuchera, R.Shahid, J.Zhao, A.Edmondson et al. - Phys.Rev. C 110, 064302 (2024).}\\
&\parbox[t]{6in}{\addtolength{\parindent}{-0.25cm} \textit{Evidence for the \ensuremath{^{\textnormal{15}}}Be ground state from \ensuremath{^{\textnormal{12}}}Be+3n events.}}\\
\end{longtable}
\end{document}